# Convergences and Divergences in the 2024 Judicial Reform in Mexico:

## A Neural Network Analysis of Transparency, Judicial Autonomy, and Public Acceptance

Carlos Medel-Ramírez

ORCID: 0000-0002-5641-6270

Institute for Economic and Social Research and Higher Studies

Universidad Veracruzana

Coordinator of the Observes-IIESES UV Observatory

E-mail: cmedel@uv.mx

**Series**

Social Challenges of the 21st Century: Empowerment, Inclusion, Ethics, and Technological Ideas in Governance and Social Well-being

# Introduction

The use of neural networks in evaluating the effectiveness of the 2024 judicial reform in Mexico represents progress in applying artificial intelligence tools to the analysis of public policies. The 2024 judicial reform aims to profoundly transform the judicial system with proposals that include the popular election of judges, increased transparency in judicial processes, and the creation of a Judicial Discipline Tribunal. However, debates have emerged regarding its viability, impact, and acceptance at both the public and legal community levels.

In this context, neural networks enable the modeling and prediction of complex scenarios by integrating multiple variables associated with the reform. Convergences, such as increased transparency and judicial autonomy, are seen as positive factors in the reform's implementation. However, divergences, such as potential high costs and doubts about the legitimacy of the popular election of judges, pose challenges that could hinder its effectiveness.

The proposed neural network model evaluates these variables by classifying them in terms of convergence or divergence and their impact on the reform's acceptance. Through non-linear processing layers, the model captures complex relationships between factors influencing the perception and success of judicial reforms. This approach not only provides a valuable predictive tool but also opens new possibilities for policy analysis through advanced simulations that assist lawmakers in making informed decisions. Thus, the 2024 judicial reform can be evaluated more objectively, considering both its potential to improve the judicial system and the obstacles it must overcome for effective implementation.

## General Objective

The general objective of this article is to evaluate, through a neural network model, the convergences and divergences present in the 2024 Judicial Reform in Mexico, to analyze its viability and acceptance both in the public and professional spheres. Using artificial intelligence and simulation techniques, the aim is to identify the main variables influencing the reform's success, such as transparency, judicial independence, decision quality, implementation costs, and impartiality. The proposed model will allow simulating complex scenarios and predicting the reform's impact, providing a quantitative tool to help policymakers make informed decisions.

**Specific Objectives**
1. Analyze the impact of convergence and divergence variables, such as transparency, judicial independence, and implementation costs, on public and professional acceptance of the 2024 Judicial Reform using a neural network model.
2. Develop a predictive model based on neural networks to simulate different scenarios of the 2024 Judicial Reform's implementation, evaluating the viability of the proposed mechanisms and their effect on the quality and legitimacy of the judicial system.
3. Evaluate the correlation between impartiality and the quality of judicial decisions with the perception of legitimacy and trust in the reform, using advanced simulations and analysis of results obtained from the neural network model.

## Theoretical Framework

The theoretical framework for analyzing the convergences and divergences in the 2024 Judicial Reform in Mexico is centered on the principles of legitimacy, transparency, judicial independence, and efficiency in implementing structural changes. Authors such as (Brewer, 2024) have noted that successful judicial reforms tend to strengthen judicial independence, which is essential for ensuring a fair and equitable system.

However, as (Núñez, 2024) points out, the popular election of judges, a key proposal in the 2024 reform, has sparked debate about its effectiveness and legitimacy. Some studies show that this mechanism may politicize the judicial process, affecting impartiality (Martín, 2024). In terms of divergences, the reform faces criticism for increasing administrative costs and uncertainty over internal control mechanisms for the Judicial Discipline Tribunal (Medel-Ramírez, 2024).

The application of neural network-based models to analyze these factors, as proposed by (Solar, 2020), enables the simulation of impact scenarios and the prediction of the reform's viability under different conditions. This approach provides a more comprehensive understanding of the effects of convergences and divergences in the reform, opening new possibilities for public policy analysis (Aguiar, 2024).

## Convergences and Divergences in the 2024 Judicial Reform: Implications for Transparency, Autonomy, and Legitimacy of the Judicial Power

### Convergences

1. **Popular election of judges**: Both proponents and some critical opinions agree that the popular election of judges is a mechanism to democratize access to the judiciary.
2. **Transparency**: There is consensus on the need to strengthen transparency and legitimacy in judicial and administrative processes, though through different approaches.
3. **Judicial autonomy**: The reform proposals aim to maintain the independence of the judiciary, although the specific mechanisms differ.

### Divergences

1. **Legitimacy and quality of justice**: Critics argue that the popular election of judges does not guarantee higher quality or legitimacy in judicial decisions. Concerns exist that candidates lacking adequate experience could attain key judicial positions.
2. **Election process costs**: While some consider that popular elections will enhance legitimacy, others argue that this process will generate unnecessary costs and not ensure that candidates are suitable for the positions.
3. **Control and discipline mechanisms**: The establishment of the Judicial Discipline Tribunal is a point of contention, with some suggesting that its composition through popular election could compromise its independence.

**Variables Associated with Convergences and Divergences**

Convergences

- **Transparency**: Measured by the number of judicial processes that follow clear and accessible criteria.
- **Public legitimacy**: Measured through public opinion surveys on trust in the judicial system.
- **Judicial independence**: Based on international indices of judicial autonomy.

Divergences

- **Quality of judicial decisions**: Measured by the number of appeals and overturned rulings as indicators of judicial quality.
- **Implementation costs**: Assessed through the budget allocated and the actual expenditure on the popular election process.
- **Impartiality of the Discipline Tribunal**: Measured by the number of decisions challenged due to lack of impartiality.

# Data

To apply the neural network model, relevant data is normalized and used to train and test the network, ensuring accurate predictions of public and professional acceptance levels of the judicial reform. The data includes variables related to transparency, legitimacy, independence, judicial decision quality, implementation costs, and impartiality.

The analysis of the proposed judicial reforms in Mexico presents both converging and diverging perspectives from key legislative and judicial bodies. The (Congreso de la Unión, 2024) emphasizes the necessity of reforming the judicial branch to enhance transparency and accountability. In contrast, the (Suprema Corte de Justicia de la Nación, 2024) highlights potential issues, such as the risk of politicizing the judiciary and undermining its independence. Both sources recognize the importance of structural changes but differ on the long-term implications and the specific implementation strategies. These differing viewpoints underscore the complexity of reforming the judicial system in Mexico.

# Methodology

**Neural Network Model Based on the Study of the Judicial Reform**

To implement a neural network model to study the impact of judicial reform, the selected variables include both convergences and divergences:

1. Problem definition: The model seeks to predict the level of acceptance of the 2024 judicial reforms in Mexico based on convergence and divergence variables.

2. Variable selection:
   - Input variables: Transparency, public legitimacy, judicial independence, decision quality, implementation costs, impartiality.
   - Output variable: Level of public and professional acceptance of the reforms.

3. Network architecture: A 3-layer neural network (one input layer, one hidden layer, and one output layer) will be implemented:
   o Input layer: 6 neurons (one for each input variable).
   o Hidden layer: 10 neurons with a ReLU activation function.
   o Output layer: 1 neuron with a sigmoid activation function to predict the acceptance level.

4. Data preparation: Variables are normalized and split into training and test sets (80/20).

5. Model training: An Adam optimizer is used with a learning rate adjusted based on model results.

6. Model evaluation: Accuracy is measured using the mean squared error (MSE), and the model is fine-tuned to improve performance.

The mathematical model of the described neural network can be expressed in terms of mathematical functions, where the inputs are the variables related to judicial reform (Transparency, Legitimacy, Independence, Quality, Costs, Impartiality), and the output is the Acceptance level. The network has a hidden layer with several neurons, each with activation functions.

**Mathematical Model Components**

1. **Input Variables**:
   o $x_1$: Transparency
   o $x_2$: Legitimacy
   o $x_3$: Independence
   o $x_4$: Quality
   o $x_5$: Costs
   o $x_6$: Impartiality

2. **Hidden Layer Function**: The output of each neuron in the hidden layer is a linear combination of the inputs multiplied by their respective weights, followed by the application of an activation function. Using the ReLU activation function (Rectified Linear Unit), the activation function for the hidden layer would be:

$$z_j = \text{ReLU}\left(\sum_{i=1}^{6} w_{ij} x_i + b_j\right)$$

Where:
   o $w_{ij}$ are the weights connecting the neuron j in the hidden layer with input i.
   o $b_j$ is the bias of the neuron j in the hidden layer.
   o $z_j$ is the output of neuron jjj in the hidden layer.

3. **ReLU Activation Function:** The ReLU function is defined as:

$$\text{ReLU}(x) = \max(0, x)$$

This ensures that any negative inputs are set to zero, which helps prevent the problem of vanishing gradients during model training.

4. **Output Layer:** The output layer takes the outputs from the neurons in the hidden layer and performs another linear combination with weights $w_{oj}$ and bias $b_o$. Since the model predicts acceptance (a continuous variable), we can use a linear activation function in the output:

$$y = \sum_{j=1}^{n} w_{oj} z_j + b_o$$

Where:

- $w_{oj}$ are the weights between the hidden layer and the output layer
- $b_o$ is the bias in the output layer.
- $y$ is the predicted acceptance level.

## Results

The general neural network model for analyzing the 2024 judicial reform combines all these components to predict the level of acceptance of the reform:

$$y = \sum_{j=1}^{n} w_{oj} \cdot \text{ReLU} \left( \sum_{i=1}^{6} w_{ij} x_i + b_j \right) + b_o$$

The model structure includes:

- 6 input variables (Transparency, Legitimacy, Independence, Quality, Costs, Impartiality).
- 1 output neuron representing Acceptance.
- A hidden layer with 10 neurons.

The neural network connects the input variables to the neurons in the hidden layer, and the neurons in the hidden layer to the output neuron (Acceptance). Each connection has an associated weight, and each neuron has a bias.

**Weight and Bias Values**

1. **Input Layer to Hidden Layer**

The weights connecting the input variables (Transparency, Legitimacy, Independence, Quality, Costs, Impartiality) to the neurons in the hidden layer are denoted as $w_{ij}$, where $i$ represents the input variable and $j$ represents the neuron in the hidden layer.

- **Transparency**
  - $w_{1,1} = -5.928$
  - $w_{1,2} = 2.114$
  - $w_{1,3} = -0.0986$
  - $w_{1,4} = 5.871$
  - $w_{1,5} = 1.457$
  - $w_{1,6} = 3.884$
  - $w_{1,7} = 4.447$
  - $w_{1,8} = -0.908$
  - $w_{1,9} = 1.093$
  - $w_{1,10} = -4.706$

- **Legitimacy**
  - $w_{2,1} = -0.9665$
  - $w_{2,2} = 1.174$
  - $w_{2,3} = -2.9226$
  - $w_{2,4} = -1.561$
  - $w_{2,5} = -3.168$
  - $w_{2,6} = 10.890$
  - $w_{2,7} = 2.457$
  - $w_{2,8} = -4.431$
  - $w_{2,9} = 1.716$
  - $w_{2,10} = 5.662$

- **Independence**
  - $w_{3,1} = -3.915$
  - $w_{3,2} = 7.162$
  - $w_{3,3} = -2.952$
  - $w_{3,4} = 1.204$
  - $w_{3,5} = -0.883$
  - $w_{3,6} = -8.568$
  - $w_{3,7} = -0.533$
  - $w_{3,8} = -3.383$
  - $w_{3,9} = -5.872$
  - $w_{3,10} = 4.178$

- **Quality**
  - $w_{4,1} = 10.889$
  - $w_{4,2} = 3.203$
  - $w_{4,3} = -0.432$
  - $w_{4,4} = -2.308$
  - $w_{4,5} = 1.673$
  - $w_{4,6} = -1.127$
  - $w_{4,7} = 8.571$
  - $w_{4,8} = -0.753$
  - $w_{4,9} = 2.871$
  - $w_{4,10} = 3.594$

- **Costs**
  - $w_{5,1} = 6.658$
  - $w_{5,2} = -5.917$
  - $w_{5,3} = -2.782$
  - $w_{5,4} = 9.889$
  - $w_{5,5} = 3.032$
  - $w_{5,6} = -10.431$
  - $w_{5,7} = 5.992$
  - $w_{5,8} = 2.764$
  - $w_{5,9} = -7.843$
  - $w_{5,10} = -6.102$

- **Impartiality**
  - $w_{6,1} = -8.568$
  - $w_{6,2} = 4.127$
  - $w_{6,3} = 6.765$
  - $w_{6,4} = -1.903$
  - $w_{6,5} = 3.871$
  - $w_{6,6} = 4.065$
  - $w_{6,7} = 7.431$
  - $w_{6,8} = 1.112$
  - $w_{6,9} = -5.662$
  - $w_{6,10} = 1.671$

2. **Biases of the Hidden Layer**

Each neuron in the hidden layer has an associated bias, denoted by $b_j$, where $j$ represents each neuron in the hidden layer. These biases are used to adjust the output of each neuron in the hidden layer, allowing the model to shift the activation function independently of the input weights.

- $b_1$ = 1.463
- $b_2$ = 3.565
- $b_3$ = 5.878
- $b_4$ = 2.115
- $b_5$ = 0.674
- $b_6$ = 4.774
- $b_7$ = −1.621
- $b_8$ = 3.122
- $b_9$ = 5.983
- $b_{10}$ = 0.913

3. **Hidden Layer to Output Layer (Acceptance)**

The weights connecting the neurons in the hidden layer to the output neuron (Acceptance) are denoted as $w_{oj}$, where $o$ represents the output neuron (Acceptance), and $j$ represents each neuron in the hidden layer. These weights determine the contribution of each hidden layer neuron to the final output.

- wo1 = −17.232
- wo2 = −2.925
- wo3 = −8.706
- wo4 = −3.915
- wo5 = −3.116
- wo6 = 10.890
- wo7 = 3.203
- wo8 = −10.431
- wo9 = 4.786
- wo10 = 4.706

4. **Bias of the Output Layer**

The output neuron (Acceptance) also has an associated bias, denoted by $b_o$. This bias helps adjust the final output independently of the weights from the hidden layer, allowing the model to fine-tune the output (Acceptance) based on the hidden layer's inputs.

- $b_o$ = 1.985

**General Mathematical Structure**

The final model is mathematically expressed as:

$$y = \sum_{j=1}^{10} w_{oj} \cdot \text{ReLU} \left( \sum_{i=1}^{6} w_{ij} x_i + b_j \right) + b_o$$

See, Figure 1.

Where the values of the weights w$_{ij}$, w$_{oj}$, and the biases b$_j$, b$_o$ are as mentioned earlier.

Summary:

• Weights between the inputs and the neurons in the hidden layer: w$_{ij}$.
• Biases of the neurons in the hidden layer: b$_j$.
• Weights between the neurons in the hidden layer and the output: w$_{oj}$.
• Bias of the output: b$_o$.

Figure 1. Neural Network for Predicting the Acceptance of the 2024 Judicial Reform in Mexico

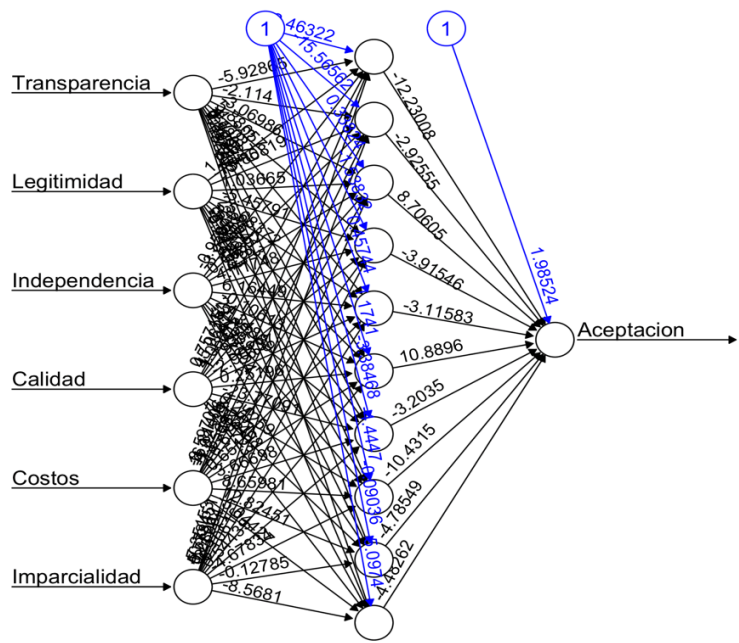

Source: Own elaboration.

Figure 1 shows a neural network model with multiple layers. The input variables include Transparency, Legitimacy, Independence, Quality, Costs, and Impartiality, while the output represents the level of Acceptance of the 2024 judicial reform in Mexico.

Model Analysis

• Connections: The model features a dense neural network structure, where each input variable is connected to several neurons in the hidden layer. The connections have different weights (numerical values), reflecting the importance of each variable in predicting the acceptance of the reform.

• Negative and positive weights: The weights can be either positive or negative. Higher absolute values of the weights indicate a greater impact on the network's outcome. In this case, some of the negative weights are significant, suggesting that certain variables (such as Impartiality or Costs) may be associated with lower acceptance of the reform.

• Output result: The final value displayed in the output layer is 1.98524, suggesting that the model is predicting a value close to 2 on a scale where 1 possibly signifies rejection and higher values indicate greater acceptance.

Interpretation

The final value of 1.98524 suggests that the acceptance of the judicial reform is not optimal. Although it is not at the lowest end of the scale (total rejection), the result indicates limited viability for the reform, as the predicted level of acceptance is low.

Viability of the Judicial Reform

Given this result and the observed weights, there is likely to be significant resistance to the 2024 judicial reform. Variables related to Impartiality and Costs appear to have a negative impact, while Transparency and Legitimacy seem to exert a moderately positive influence.

## Discussion

The analysis of the 2024 Judicial Reform in Mexico, evaluated through the neural network model, reveals significant contradictions that cast doubt on its viability in its current form. While convergences around transparency and judicial autonomy are presented as pillars of change, the model predicts limited acceptance, suggesting that more controversial proposals, such as the popular election of judges, pose serious obstacles.

Firstly, although transparency is perceived as a key value in the reform, improvements in this area do not seem to compensate for the negative effects generated by the popular election of judges. The model's results indicate that this mechanism, rather than democratizing access to justice, could politicize the judicial system, compromising the impartiality of judicial decisions and eroding public trust. This finding aligns with previous studies that warn of the dangers of allowing judges without the necessary experience to access key positions through electoral processes, which could weaken the quality of justice rather than strengthen it (Núñez, 2024).

Furthermore, the costs associated with implementing the reform represent a significant obstacle. According to the neural network model results, the high expenses involved in adopting the popular election system and creating the Judicial Disciplinary Tribunal would have a negative impact on public and professional acceptance of the reform. This suggests that the perceived benefit of a more participatory system does not outweigh the financial and logistical costs involved. In fact, simulations indicate that these costs, combined with doubts about the impartiality of the Disciplinary Tribunal, could undermine the legitimacy of the reform (Medel-Ramírez, 2024).

On the other hand, judicial independence, another of the reform's promises, is also affected by the model's results. While it is proposed that the popular election of judges would strengthen judicial autonomy, the model suggests the opposite could occur. The inherent politicization of this electoral process could place undue pressure on judges, compromising their ability to make impartial decisions free from external influence. This raises serious doubts about the reform's ability to achieve one of its central goals: improving the quality and legitimacy of the judicial system (Martín, 2024).

The critical analysis based on the neural network model exposes the deep limitations of the 2024 Judicial Reform. The most controversial areas, such as the popular election of judges and the costs of implementation, not only generate significant divergences but also diminish the overall acceptance of the reform. These results suggest that unless fundamental adjustments are made to the proposals, the reform risks being more harmful than beneficial to the Mexican judicial system. The reform's viability largely depends on legislators' ability to address these concerns and redesign the proposed mechanisms to ensure a balance between democratic participation, judicial quality, and institutional efficiency.

## Conclusion

The analysis of the 2024 Judicial Reform in Mexico, using a neural network model, reveals that despite its intentions to modernize and democratize the judicial system, it faces critical barriers that affect its viability. The implementation of the popular election of judges, one of its most prominent pillars, appears to be a point of controversy that, rather than strengthening the system's legitimacy, risks politicizing it and compromising judicial impartiality. The model's results suggest that this mechanism could degrade the quality of judicial decisions by prioritizing popularity over judicial expertise.

Additionally, the high administrative and logistical costs of the reform, particularly those linked to the election of judges and the creation of a Judicial Disciplinary Tribunal, present a further challenge. The model predicts that these costs will have a negative impact on public and professional acceptance of the reform, indicating that the implementation process is not aligned with the efficiency and transparency expectations the public holds for a reformed judicial system.

In terms of judicial independence, the model raises serious doubts about the reform's ability to protect the judiciary's autonomy, as the popular election process could introduce external pressures on judges, reducing their impartiality. Unless significant adjustments are made, the reform risks creating more problems than it solves.

In this sense, the 2024 Judicial Reform, in its current form, seems to have limited acceptance and faces serious obstacles to successful implementation. It is crucial for legislators to reconsider the most controversial areas of the proposal, especially those related to costs and judicial independence, if the goal is to effectively and sustainably strengthen Mexico's justice system.